# Geometrical complexity of the antidots unit cell effect on the spin wave excitations spectra

M. Zelent,[1] N. Tahir[2,4], R. Gieniusz[2], J. W. Kłos[1], T. Wojciechowski[2], U. Guzowska[2], A. Maziewski[2], J. Ding[3], A.O. Adeyeye[3] and M. Krawczyk[1]

[1]*Faculty of Physics, Adam Mickiewicz University in Poznan, Umultowska 85, 61–614 Poznań, Poland*
[2]*Faculty of Physics, University of Bialystok, ul. Konstantego Ciołkowskiego 1L, Białystok 15–245, Poland*
[3]*Department of Electrical and Computer Engineering, National University of Singapore, Singapore*
[4]*Division of Science and Technology, University of education, Lahore, Pakistan*

**Abstract:**

We have investigated theoretically (with micromagnetic simulations and plane wave method) and experimentally (with ferromagnetic resonance and Brillouin light scattering) three types of antidot lattices (ADLs) based on permalloy thin films with increased complexity of the unit cell: simple square, bi-component square and wave-like ADL. We have found that placing a small additional antidot in the center of the unit cell of the square ADL modify significantly the spin wave spectrum and its dependence on the orientation of the in-plane magnetic field. We also check the further changes of spin wave spectrum resulting from the introduction of air-gaps connecting small and large antidots. In particular, the presence of small antidots change the dependence of the frequency of the fundamental mode on the angle of the in-plane applied magnetic field. The air-gaps strongly discriminates the propagation of spin waves in two principal direction of ADL lattice, orthogonal to each other. In spite of these spectral changes, the spatial distribution of the spin wave amplitude generally preserves some similarities for all three structures. We also highlighted out the role of defects in the ADL in the observed spectra. The obtained results can be interesting for the magnonics applications of the magnonic crystals.

PACS: 75.30.Ds, 75.40.Gb, 76.50.+g, 75.78.Cd

## I.   Introduction

Recent studies have shown that spin waves (SWs) in nano-patterned magnetic media can be used for signal processing, communication and sensing.[1,2,3] The information can be encoded in the amplitude[4], phase[5] or frequency of SWs, enabling the design of a new class of wave-based information, sensing and communications technology devices. Compared to conventional microwave devices, those based on SWs can be orders of magnitude smaller, as SWs operating at GHz or THz frequencies have micrometric or nanometric wavelength, i.e. much shorter than wavelength of corresponding electromagnetic waves at the same frequency.[6] In reference to electronic devices, the devices based on SWs can operate in much higher frequencies and consume two orders less energy. These prospects stimulated the emerging research fields of magnonics and magnon- spintronics.[6,7]

The important class of the magnonic structures are patterned thin films, especially these with the artificially introduced periodic pattern – magnonic crystals (MCs).[7,8,9] The SW spectrum in MCs is characterized by frequency bands and forbidden frequency band gaps whose position can be easily tuned by a magnetic field magnitude or its orientation.[10,11,12] Moreover, also other peculiar properties, like *nonreciprocity* of SWs[13,14,15] (characterized by asymmetric amplitude distribution and/or asymmetric dispersion relation: $\omega(\boldsymbol{k}) \neq \omega(-\boldsymbol{k})$, where $\omega$ is the angular frequency and *k* is a wave vector) or *re-programmability*[10,16] (where by change of the static magnetic configuration we can control the SW dispersion) are interesting from fundamental reasons. These unique features of MCs can be also potentially useful for applications required extended functionalities over standard ones offered by photonic or phononic structures.

Most of the devices proposed in the past consisted of one- or two-dimensional (1D and 2D) yttrium iron garnet (YIG) films with an artificial micron-size periodicity and SWs generated by inductive methods.[17] YIG has the advantage of being a very low-damping material, but its integration into standard semiconductor technology is challenging due to the difficulties of growing and patterning high quality thin films. Recently, fabrication of ultrathin YIG films with pulsed layer deposition technique was established.[18] However, their integration with CMOS technology still remains challenging.[19] Polycrystalline magnetic alloys such as NiFe (permalloy, Py) or CoFeB, are better suited to meet the industrial demand of integration and miniaturization, but their drawback is the relatively high SW damping which prevents propagation above a few microns. This can be compensated by downscaling the MCs to the nanometric size, also thanks to the new possibility of launching SWs with sub-micrometric wavelength by spin-transfer torque,[20] spin Hall effect[21] or microwave transducers used over periodically patterned substrates – 2D MCs.[22] Therefore, 2D MCs obtained from metallic alloys are considered for the next generation of the above mentioned applications.[23,24,25,26] Particular interest is in the 2D antidots lattices (ADLs) MCs,[27] where the periodic array of holes is drilled in the ferromagnetic film. It results from relative simplicity of the fabrication. The ADLs can be considered as a network of multiple connected magnonic waveguides having an effective width equal to the distance between neighboring hole edges, where both confined and propagating SWs coexist.[28,29,30,31] Further modification of the SW spectra can be achieved by filling the antidots with a ferromagnetic material in direct contact or separated from the host material.[32,33,34,35] Despite the numerous studies, the comprehension of the relationship between the geometric characteristics of the MCs unit cell, especially in ADLs and the SWs dynamic properties still needs to be developed.

In the present work, we selected the square array of cylindrical holes (square ADL – Fig. 1(a)) in thin Py film as a base structure. Then we complicated the structure by placing an additional smaller hole at the center of the unit cell (bi-component ADL – Fig.1b). The results from these two reference structures we used to interpret the spectrum obtained for the wave-like ADL. The wave-like ADL structure is derived from bi-component ADL by drilling the narrow connections between nearest larger and smaller holes along the one principal direction of ADL– horizontal axis in Fig. 1(c). The square ADLs were broadly investigated in the past, also with various shapes of the antidots.[36] However, there are only few research papers reporting investigations of the SW dynamics in thin film bi-component ADLs[37,38,39] moreover, for wave-like structures the detailed scientific reports are still missing. From the previous studies,[40] we know, that these modifications of the square unit cell introduce significant changes in the magnetization reversal anisotropy map, making them interesting for data storage applications and for investigation of the SW dynamics. We use ferromagnetic resonance (FMR) and Brillouin light scattering (BLS) techniques to study standing and propagating SWs in this set of ADLs. We found that the sensitivity of the FMR spectra on orientation of the magnetic field is strongly dependent on the minor changes of the antidot pattern in the unit cell, even if the unit cell preserves the same symmetry. The strongest angular dependence of FMR spectra has been found for the wave-like pattern. For this structure the changes of the demagnetizing field induces significant changes in the SWs spectrum. Here, the rotation of magnetic field from the direction parallel to the wave-like pattern transforms fundamental SW excitation into the edge modes, which leads to the loses of their intensity in BLS spectra. Also the dispersion relations of SWs are strongly affected by the modification of antidot pattern in the unit cell, even if this modification cover less than few percent of the unit cell area. Our experimental results are successfully interpreted on the basis of calculations performed with the plane wave method (PWM) [41] and micromagnetic simulations (MS). Moreover, the comparison of experimental and numerical results allows us also to identify influence of distortions of the antidots shape and the lattice defects on the FMR spectra. Thus, a general view that governs the SW characteristics in square lattice ADLs with complex unit cell has been presented. The delivered knowledge will be useful for design of magnonic devices operating at the microwave frequency range.

The paper is organized as follows. In the Sec. II we present the fabrication process of the ADLs and experimental methods used in investigations of the SW dynamics. In Sec. III we present results of the FMR measurements, which are interpreted with the results of the MSs. The propagative properties of SWs are discussed in Sec. IV, where the results of BLS measurements and PWM calculations are exploited. In the last section we summaries our results.

## II. Fabrication and experimental methods

Large area (4 × 4 mm$^2$) of $Ni_{80}Fe_{20}$ (Py) antidot nanostructures were patterned on commercially available silicon (Si) substrates by employing deep ultraviolet lithography (DUV) at 248 nm exposing wavelength. In order to create patterns in the resist, the substrate was spin coated with 60 nm thick bottom anti-reflecting coating (BARC) followed by a 480 nm positive deep UV-photoresist which is four to five time thicker than those typically used for e-beam lithography. Thicker resist helps to achieve high aspect ratio and additionally makes the lift-off process easier. A Nikon lithographic scanner with KrF excimer laser radiation was used to expose the resist. In order to transfer resist patterns into antidots, 10 nm thick Py was deposited at room temperature by e-beam evaporation technique at rate of 0.2 Å/s while the pressure in the chamber was maintained at $2 \times 10^{-6}$ Torr.

Showed in Fig. 1, the samples with various geometries starting from basic square ADL (a), to the most complex system (wave-like pattern, (c)) by the systematically inclusion of additional antidots (bi-component ADL with alternating diameters of circular antidots (b)) were fabricated by the procedure discussed above. In order to keep the edges of the structures sharp, the last step of the sample fabrication process (lift-off of BARC) was not carried out so BARC is present inside the antidots lattice which is confirmed by scanning electron microscope (SEM). The antidots are close to the circular shape, however the larger deformations are visible for a square and wave-like ADLs (see insets in Fig. 1 (a) and (c)). In square ADL the diameter of antidots $D$ is 405 nm and period $a$ = 650 nm, in bi-component ADL the diameter of larger and smaller antidots is 410 and 140 nm, respectively, and period 650 nm. In the wave-like structure larger and smaller antidots are 350 and 130 nm, and period is 580 nm. We introduce a coordinating system with the $x$-axis along the horizontal direction in Fig. 1, connecting next-nearest larger antidots. Structural analysis was carried out by ZEISS EVO HD15 scanning electron microscope (SEM).

Dynamics of the magnetization was studied through conventional X-band FMR at constant frequency 9.4 GHz in dependence on the static magnetic field $H$ orientation in the film plane. The orientation of the $H$ field is described by the angle $\phi_H$, where $\phi_H$ = 0 is a direction along the $x$-axis. The SW dispersion relations were measured with BLS. In these measurements we limit our studies to the Damon-Eshbach (DE) geometry (propagation of SWs perpendicular to the direction of the magnetization saturation). At this configuration the larger group velocity of SWs is expected. All BLS measurements were performed at room temperature in backscattering geometry using Sandercock (3+3)-pass tandem Fabry-Perot interferometer to analyze the frequency shift of light from a single-mode solid state laser with λ = 532 nm.[42,43] Inelastically scattered light was sent through a crossed analyzer in order to suppress surface phonons signal. A static magnetic field was applied in the film-plane perpendicular to the transferred wave vector scattering plane. The measurements were performed for various magnetic field values and at different angles of incidence of the probing light beam, i.e. the angle between the direction of the incident laser beam and the film normal θ. The amplitude of the SW in-plane wave vector $q$ is related to the angle of incidence by the relation $q = (4\pi/\lambda)\sin\theta$. By changing θ in the range of 10°-80°, it is possible to change of $q$ in the range of $0.81 \times 10^5 - 1.8 \times 10^5$ cm$^{-1}$.

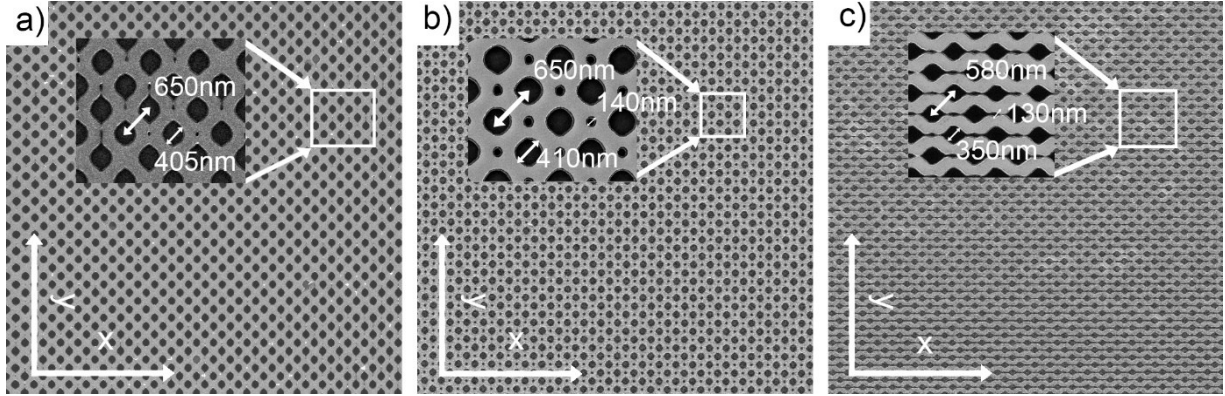

**Fig. 1.** SEM images of the three samplesbased on Py film investigated in the paper: a) square ADL, b) bi-component ADL – composed to two kind of antidots and c) wave-like ALD, where lager and smaller antidots are connected by air tranches. The thickness of Py layer is 10 nm.

## III. FMR spectra

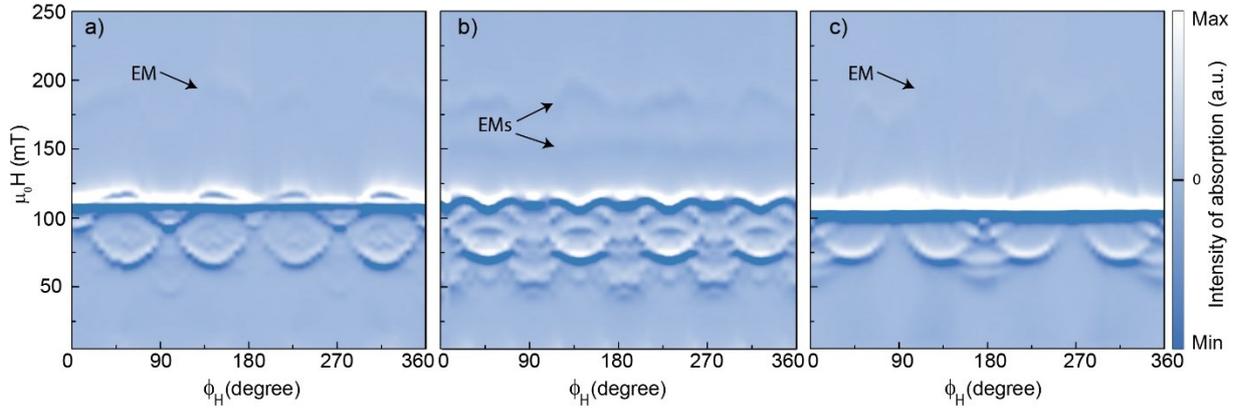

Fig. 2. FMR spectra in dependence on the in-plane angle of the magnetic field with respect to the *x*-axis for the ADL samples presented in Fig. 1: (a) square ADL, (b) bi-component ADL and (c) wave-like ADL. The intensity of absorption of the FMR signal is shown in color scale. The color labeled by 0 marks the FMR resonance field. The labels EM point weakly visible edge modes.

**Measurements.** The FMR spectra in dependence on the static magnetic field orientation are shown in Fig. 2. The results for the patterned samples can be compared with the data for the uniform Py film of 10 nm thickness, being a reference sample. In FMR spectra of the reference sample, a single resonance line of constant position (i.e. independent on the angle $\phi_H$) is observed at 102 mT. This line corresponds to the resonance of a uniform mode. The lack of any angular dependence for this resonance confirms that the reference sample has negligible in-plane anisotropy.[40]

In every FMR spectra of the ADL we see the single line of very high intensity, whose resonant magnetic field is close to the FMR field of the reference sample. We correlate this line with an excitation of the fundamental mode (FM). This mode has in-phase SW oscillations in the whole unit cell. Apart from the FM line, there are lines with the resonance fields lower and higher than FM field. We attribute these lines to the bulk modes (BMs), with oscillating amplitude of the SW in the bulk regions of ADL, and the edge modes (EMs) with SW amplitude concentrated at edges of the antidots. The deeper insight into above discussed modes will be presented latter with the aid results obtained with MS. Now, we will discuss the angular variation of FMR lines of the FM and BM modes in all samples, after that for the EMs.

Fig. 2 (a) shows a 2D color map of the FMR spectra collected at different magnetic field orientations in the square ADL. The presence of a four-fold symmetry in the spectra can be recognized.[44,45,46] However, there are also some noticeable details (at low magnetic fields) which discriminate 0 and 90º field directions, nominally equivalent orientations. We attribute these features of FMR spectrum to the deviation of the antidots from a circular shape seen in the SEM image in Fig. 1(a). For various in-plane angles, we can separate two clear resonances of the BMs, one around 93 mT at smaller angles ($\phi_H < 20º$) and for larger angles another BM becomes intensive, with resonant field decreasing to 69 mT at $\phi_H = 45º$. For both of them the variation of its magnetic field position with the change of the magnetic field orientation is observed. We also notice, that the most intensive resonance line doesn't depend on the magnetic field orientation, which is unexpected results for ADLs.[46,47] The cause of such independence will be discussed later, after description of the data of the wave-like patterns (presented in Fig. 2c), where similar effect is observed.

Fig. 2(b) presents the FMR results for the bi-component ADL. The slight oscillation of the FM resonant field with an angle $\phi_H$ is clear and the FMR spectra at $\phi_H = 90º$ is almost the same as at $\phi_H = 0º$. These measurements confirm observation from the SEM in Fig. 1 (b), that the FMR spectra fully preserves the four fold symmetry of the considered square ADL. Plenty of BMs excited at intermediated angles are also found, with the most pronounced line around $\phi_H = 45º$ at field 73 mT. For this angle, the applied field is parallel to the side of the square lattice of antidots and this intensive excitation is close to BM measured in the square ADL (Fig. 2(a)), although signal is more intensive in bi-component ADL (comparable to the intensity of the FM). This points at the origin of this excitation from the larger antidots. Additional BM resonances of low intensities, not found in the square ADL could be attributed to the presence of smaller antidots. Nevertheless, the spectra of square and bi-component ADL are qualitatively different for other angles. For the $\phi_H = 0º$ the three lines are clearly observed at fields 42, 51, and 78 mT, while in the case of square ADL the only one resonance line apart FM was observed. At the angle $\phi_H = 18º$ the most reach spectra for bi-component ADL is detected with several resonance peaks below FM line (72, 75, 83, and 96 mT).

Fig. 2(c) presents the results of the FMR measurements for the wave-like ADL. Here, $\phi_H = 0º$ defines an angle in which external magnetic field is applied parallel to the wave-like pattern. We found two-fold symmetry axis according with the symmetry of the structure. For small $\phi_H$ resonance lines of small intensity are observed only below FM field (clearly visible on the logarithmic scale used in the color map in Fig. 2(c)). At $\phi_H = 0º$ multiple, smaller intensity resonances of BMs are found with the most intensive at 60, 70 and 90 mT. Around $\phi_H = 45º$ one resonance peak of significant intensity appears with minimum at 74 mT. This peak is much similar to the observed for square and bi-component ADLs (Fig. 2(a) and (b)). At $\phi_H = 90°$ the BMs are not detected. The most intensive resonance field does not change with the rotation of the external magnetic field, like in the square ADL. This is also unexpected result because the demagnetizing field changes significantly between the two orthogonal magnetic field orientations in the wave-like ADL. We attribute this insensitivity on the field orientation to the presence of structural defects. The defects present in both structures can hide the angular variation of the FM in the measured FMR spectra.[48]. Defects can been seen in Fig. 1(a) and (c) for square and wave-like ADL, respectively. Nevertheless, the variation of the EMs resonance fields are still observable, as will be discussed below.

In the square ADL (Fig. 2(a)) only a single EM is found. With change of the in-plane angle $\phi_H$ from 0 to 45° the resonance field of the EM increases. For further increase of the angle from 45° to 90° the resonance field of the EM decreases. When we compare the spectra taken at 90° and 0°, nominally equivalent directions, we note significant differences – the EM in the former (162 mT) is at significantly lower resonance field than for 0° (182 mT) and it has much smaller intensity. This points at a non-circular shape of the antidots and different curvature of their edges along orthogonal directions,[49] which indeed can be seen also in the SEM image shown in Fig. 1(a). However, in the observed dependence of the resonant field on a magnetic field orientation influence of varied static demagnetizing field due to change a distance to nearest antidots with a change of the $\phi_H$ shall be taken into account. The edge to edge separation decreases from approximately 448 to 200 nm for $\phi_H$ changed from 0 to 45° and the demagnetizing field value increases, which pushes the resonance field of the SW excitations to higher values.

There are two EMs in the bi-component ADL (Fig. 2(b)). The resonance at higher field (varied from 181 to 186 mT for $\phi_H$ changed from 0° to 44°, respectively) we attribute to the EM at the larger antidots, because at similar fields EM was observed for the square ADL (Fig. 2(a), with resonance line varied from 182 to 191 mT for respective field orientations). The resonance line at lower field (around 158 mT) is weakly dependent on the field orientation and can be connected with EM localized at smaller antidots. There is much better agreement between spectra at $\phi_H = 0$ and 90° for bi-component ADL (Fig. 2(b)), which points that the shape of antidots in this structure is close to circular (Fig. 1(b)).

For wave-like ADL we have not found the EM for the magnetic field oriented along wave-like lines (Fig. 2(c)). It detaches from the FM at $\phi_H$ around 18° and increases field with increasing $\phi_H$. The maximal resonance field is reached at $\phi_H = 45°$ (197 mT) with further increase of $\phi_H$ the resonance field of this mode slightly decreases. We note, that at around $\phi_H = 45°$ field orientation, another EM detaches from the main line, its resonant field increases with $\phi_H$ and merge with the first EM at $\phi_H = 90°$ (179 mT). The most intensive signal from the EM is detected at $H$ oriented along the $y$-axis ($\phi_H = 90°$).

**Micromagnetic simulations.** The experimental results of the FMR measurements are further analyzed through the MS with MuMax3 environment.[50] The Landau-Lifshitz (LL) equation including exchange and dipolar fields and small damping is solved with finite difference method in time and real space domain. MSs were performed for following structural parameters, kept fixed for considered structures: the thickness - 10 nm, the diameter of the larger antidots - 400 nm and the lattice constant - 650 nm. The diameter of smaller antidots in bi-component and wave-like ADLs was assumed to be 140 nm. For wave-like ADL the width of air-gaps between wave-like channels have varied (with minimum around 90 nm). The assumed lattice constant and the size of larger antidots are only slightly different as compared to the square and bi-component ADL real samples (see Fig.1). However, the diameter of larger antidots taken for MS overestimates the a diameter of this antidot in the experimental wave-like sample. We decide to keep the same size of the unit cell and the geometrical features common for all ADLs to clearly elucidate the role of introduced complexities of the unit cell. The magnetic parameters used for MSs are standard values of the Py: saturation magnetization $M_s$ = 800 kAm$^{-1}$, exchange constant $A$ = 13×10$^{-12}$ Jm$^{-1}$. A magnetocrystalline anisotropy is neglected in our calculations. Small damping constant was assumed in simulations α = 0.001. For all structures the mesh of size 2 nm × 2 nm × 10 nm was used along the x, y and z-axis respectively. Periodic boundary conditions for x- and y-direction (with 16 repetitions square cells in the supercell) are used to reduce an influence of the external ends of the structure. We employed the built-in function for edge smoothing to overcome staircase effects at the antidots edges. For excitation of the SW precession we used microwave external magnetic field in the form of *sinc* function in time domain and spatially homogeneous in the whole sample with maximal amplitude of 5.413 × 10$^{-2}$ T and cutoff frequency $f$ = 45 GHz. After collecting data through 30 ns, we performed fast Fourier transform (FFT) of the signal to get the frequency spectra of SW excitations, which are directly related to the resonance magnetic field spectra collected in FMR measurements. The relation between resonance frequencies and resonance magnetic fields are discussed in the Appendix A.

Figs. 3, 4 and 5 show in panels (a) the calculated maps of the FMR frequency spectra in dependence on $\phi_H$ (the magnetic field magnitude was fixed to $\mu_0 H$ = 0.1 T in these simulations) for square ADL, bi-component ADL and wave-like ADL, respectively. For all structures the satellite peaks are observed on both sides of the most intensive line corresponding to FM, similarly to the experimental data. To identify SW excitations, we show (in parts (b) – (g) of Fig. 3-5) the spatial distributions of SWs amplitude related to the most intensive lines (modes) in the FMR spectra. These spatial profiles of SWs were plotted for two orientation of the magnetic fields, $\phi_H$ equals to 0 and 44° in square ADL and bi-component ADL. For wave-like ADL we analyzed the profiles of SWs for a few other angles $\phi_H$. Overall, the calculated maps of the angular dependences are qualitatively in good agreement with the experimental results presented in Fig. 2. Note however, that a part of experimental spectra for low values of magnetic resonance field corresponds to the high frequency part of the simulated spectra (see Appendix A for clarification).

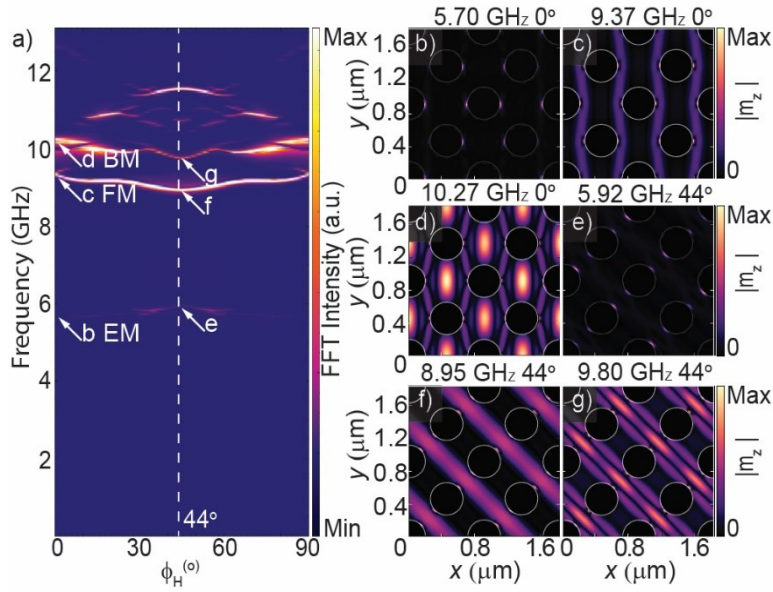

**Fig. 3.** (a) FMR frequency spectra of the square ADL obtained from MSs in dependence on the in-plane angle of the external magnetic field with respect to the *x*-axis ($\mu_0 H = 0.1$ T). (b) Amplitude of the SWs marked with arrows in (a) for $\phi_H = 0°$ (along the *x*-axis) and $\phi_H = 44°$.

In the square ADL a single low frequency mode is recognized as an EM with in-phase oscillations on both sides of the antidots (see Fig. 3(b) and (e)). Its frequency slightly varies from 5.7 to 5.92 GHz in the range of $\phi_H$ from 0 to 44°. This frequency variation of the EM is solely due to demagnetizing field which depends on $\phi_H$. In Fig. 3(c) and (f) we show the spatial profile of the FM. The changes of frequencies of FM in dependence on $\phi_H$ correspond to the most intensive line which decreases from 9.37 GHz at $\phi_H = 0$ to 8.95 GHz at $\phi_H = 45°$. This confirms, that independence on $\phi_H$ of the resonance magnetic field found in measurements (Fig. 2(a)) does not exist in an ideal structure investigated in MSs, pointing out that the experimental result is an effect of the irregularities in the sample. Also plenty of high frequency BMs are recognized in MS spectra, with the most intensive at 10.27 GHz ($\phi_H = 0$, Fig 3(d)) and 9.80 GHz ($\phi_H = 44°$, Fig 3(g)), which have maximal amplitude concentrated between antidots along direction perpendicular to *H* and two nodal lines in the unit cell along the same direction. Its position is shifted by half of ADL period, *a*/2 as compared to the FM. We can link this mode with the line at 93 mT in the measurements shown in Fig. 2(a).

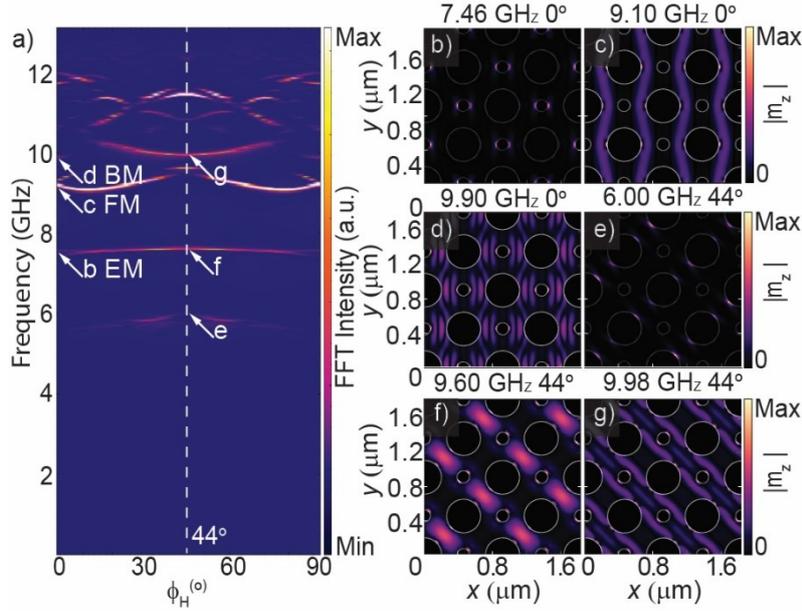

**Fig. 4.** a) Calculated FMR frequency spectra of the bi-component ADL in dependence on the in-plane angle of the magnetic field ($\mu_0 H = 0.1$ T). Visualization of the SW modes amplitude distribution with the most intensive resonances is shown in (b) for $\phi_H = 0°$ and (c) $\phi_H = 44°$ for modes marked with arrows in (a).

In Fig. 4 the results of MSs for bi-component ADL are shown. Comparing Fig. 3(a) and 4(a) one can see that additional small antidot inside the square unit cell leads to changes in the calculated spectra which overall agree with the measurements (Fig. 2(a) and (b)). The EM found in the square ADL (Fig. 3(e)) remains almost on the same frequency level in the bi-component ADL (Fig. 4(e)) and its amplitude distribution is also similar. There is additional resonance line above this EM and below FM. That new EM mode is observed at frequency 7.46 GHz ($\phi_H = 0°$) according with expectations has amplitude oscillating in-phase located near small antidots - see Fig. 4(b). The frequency of FM in bi-component ADL at $\phi_H = 0°$ is at 9.1 GHz, which is 320 MHz lower than for the square ADL. Moreover, opposite to square ADL, the FM frequency increases to 9.64 GHz (after slight decrease) with increasing $\phi_H$ for field directed alongside of the square unit cell. Both effects comes from the small holes, at 0° (Fig. 4(c)) they introduce additional demagnetizing field in the area of the FM confinement (profile very similar to the square ADL, Fig. 3(c)), while at 44° (Fig. 4(f)) they introduce quantization of the FM along the direction perpendicular to $H$. FMR frequency spectra shows also several excitations above the FM frequency with comparable intensity. These BMs show quantization along or perpendicular to the field direction. Their profiles are modified by the presence of the small antidots as compared to square ADL and their frequencies are changed as well as.

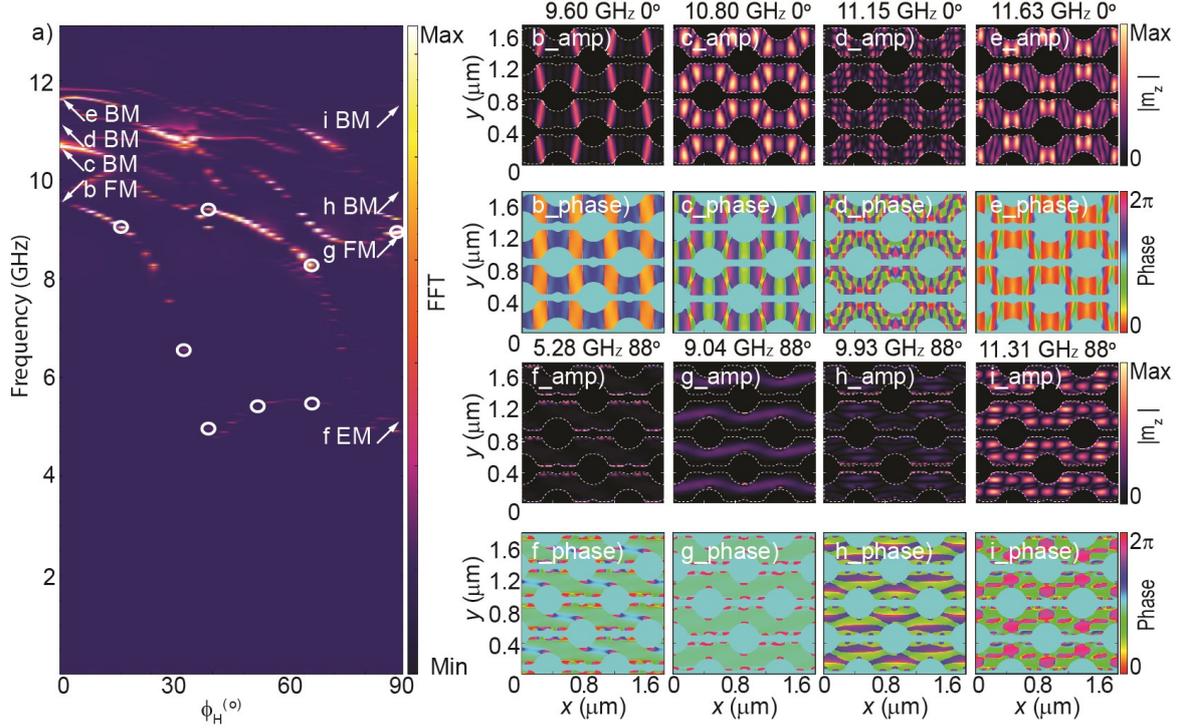

Fig. 5. a) Calculated FMR frequency spectra in dependence on the angle of the in-plane magnetic field ($\mu_0 H = 0.1$ T) for wave-like ADL. Visualization of the SW modes related to the most intensive resonances in FMR spectra for $\phi_H = 0°$ (b-e) and $\phi_H = 88°$ (f-i). In the top (bottom) row the amplitude (phase) of SWs is shown. White circles in (a) mark frequencies the FMs and EMs for which the profiles for oblique orientation of external magnetic field are presented in Fig. 6.

For the wave-like ADL the calculated FMR frequency spectrum along with the amplitude and phase distribution of exemplary SW excitations are collected in Fig. 5. We can observe drastic difference between this spectra and the spectra of the square and bi-component ADLs with interesting dependences of SWs frequencies on $\phi_H$ found. For $\phi_H = 0°$ the magnetic field and static magnetization are along the wave-like channels parallel to the *x*-axis. Even in saturation state a demagnetizing field is too weak to catch EMs, and this mode has not been found either in MSs or FMR measurements. The FM frequency (9.6 GHz) is higher than in the square and bi-component ADL, as well as in homogeneous Py film (9.3 GHz at the same field 0.1 T). Such behavior confirms the negligible impact of demagnetizing field. The amplitude of the FM mode is concentrated in the parts of the wave-like structure rotated by $\pm 45°$ to the *x*-axis and oscillate in the same phase. Nevertheless, this mode (Fig. 5(b)) still slightly reminds FM modes from Figs. 3(b) and 4(b), although it is formed in wave-like channels separated by air gaps, thus the SW excitations in neighboring wave-like channels are coupled by dipolar interactions only.

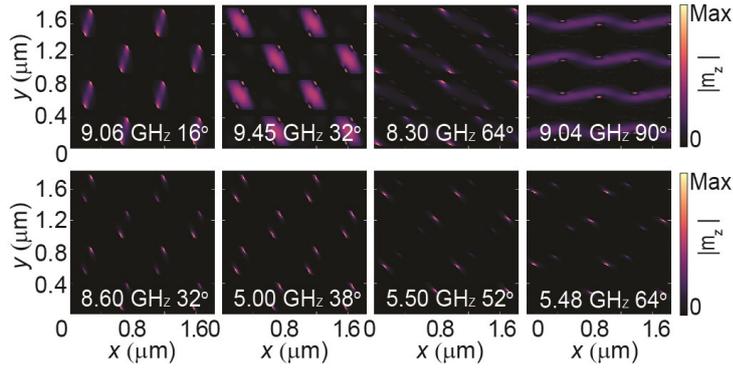

**Fig. 6.** Visualization of the SW's FM (top row) and EM (bottom row) for oblique directions of the external magnetic field in the wave-like ADL. The frequencies of these modes are marked in Fig. 5 by white circles.

During the rotation of the magnetic field from 0° to 45° the demagnetizing field decreases in the parts of the wave-like structure rotated by -45°, confining the SW amplitude in those areas and continuously transforming FM into EM. The EM observed for larger values of $\phi_H$ are concentrated at the edges of the big antidots (see transformation shown in Fig. 5 and Fig. 6 starting from b) through points marked by white circles). This mode reaches minimal frequency (~ 5 GHz) at $\phi_H = 45°$. This scenario is consistent with the experimental observations (Fig. 2(c)), where EM was observed only for $\phi_H > 18°$ and took the maximal values of the resonant field at $\phi_H = 45°$ and 135°. In measurements, the transformation of the FM is not clear, because the intensive and independent on $\phi_H$ was observed there (Fig. 2(c)). Close to $\phi_H = 30°$, when the FM mode transforms into the EM and loses its intensity, another mode appears around 9 GHz and acquires large intensity. The profile of this new FM is shown in the top row in Fig. 6 (9.45 GHz at $\phi_H = 32°$) and its amplitude distribution is very close to the profile of the FM in the bi-component ADL (Fig. 4(f)). However, the frequency of this mode in the wave-like ADL decreases (well below 7 GHz) with further increase of $\phi_H$ and its profile transforms continuously from like an FM (still at 64°, 8.30 GHz in Fig. 6) into the EM with amplitude concentrated at the edges of narrow air gaps (see in Fig. 5(f) the EM at $\phi_H = 88°$). With this increase of the $\phi_H$ the intensity of new FM decreases. Than, around 65° another one mode has higher intensity (see the profile at 88° shown in Fig. 5(g)). The profile of this mode is very close to the FM found in square and bi-component ADL (see Fig. 3(c) and 4(c)). The described scenario points out the possibility of designing a structure in which clockwise and counter-clockwise rotation of the external magnetic field will give different microwave frequency respond. This can be achieved for instance by changing thickness or width of the wave-like parts oriented at $+45^o$ and $-45^o$ with respect to the *x*-axis.

We note also that at $\phi_H = 0°$ a few BMs of high intensity are found in wave-like ADL. Their intensity decrease with rotation of the magnetic field, according with the measurements. From the profiles shown in Fig. 5(c-e) and (h-f) we can connected them to the BMs, which are quantized along field direction and concentrated in different parts of the structure.

## IV. Dispersion relation of SWs

In order to study influence of increased complexity of the unit cell on propagating properties of SWs we measured dispersion relation with the aid of BLS. To interpret experimental data the BLS cross-section, dispersion relation and the profiles of SWs (spatial distribution of the out-of-plane dynamical component of the magnetization) for the most intensive modes were computed with PWM. The spatial profiles give us the information about concentration of the SW amplitude in the unit cell. We compared thus outcomes with the results of MS for self-verification. With these information we can link the SW modes with their dynamical properties to predict the confined or propagative character of the modes. The PWM is a useful method for calculation of the SW spectra in MCs,[51] enabling calculation of the dispersion relation and profiles of the SWs for MC of any lattice type and for arbitrary shape of the elements. The Landau-Lifshitz equation is linearized, then it is transformed in to algebraic eigenproblem. The solutions in frequency domain and reciprocal space are found numerically. We use the same geometrical parameters of the ADLs as in MSs. The method is described in details in the Appendix B.

In Fig. 6(a) we present the exemplary BLS spectra for the square ADL at fixed, but nonzero, magnitude of the wave number $q = 1.8 \times 10^5$ cm$^{-1}$ corresponding to the incidence angle $\theta = 30°$. Each spectrum was collected for different values of the static magnetic field (39, 58, 76 and 96 mT) oriented in the film plane, along the *x*-axis of the square ADL. We can observe increase of the frequencies with increasing strength of the external magnetic field, which confirms the magnetic origin of the considered excitations – identified as SWs. The positions of the two most intensive peaks were marked in Fig. 6(a) by dashed lines. This allows to trace the shift of the frequencies of selected modes in dependence on strength of the magnetic field. These two modes of the highest intensities were found also in FMR spectrum (Fig. 2(a)) and identified in MSs (see Fig. 3) which refer to the Brillouin zone (BZ) center ($q = 0$). The PWM calculations were performed for the magnetic field 76 mT (related to the bolded curve of BLS spectrum in Fig. 6(a)) to keep the saturation state in the all samples, which is an important assumption of the PWM.

In MS and we found the mode of the frequency slightly below FM (see Fig. 3(b)) to be the EM. However, in PWM calculation (see Fig. 6(b)) the first SW excitation is at 7.59 GHz (at BZ center) with amplitude distribution presented in Fig. 6(e). In PWM this mode has mixed character, combining FM (amplitude concentrated in the channels perpendicular to *H*) and EM (strong amplitude at the edges of antidots oscillating in-phase with the FM). This is a result of the assumptions used PWM which adapt this method for the calculation of dispersion relation for MCs with non-magnetic materials. In the PWM the artificial material is introduced instead of the non-magnetic, which results in strong artificial pinning of the magnetization at the antidot edges. This shifts up frequency of the EM, while the other modes are almost unaffected.[51,58] For small ratio of antidots separation to thickness of the ADL,[27] like in the samples considered in this paper, the FM and EM can merge to form collective excitation shown in Fig. 6(e). Below we will consider FM and BM modes in the square ADL.

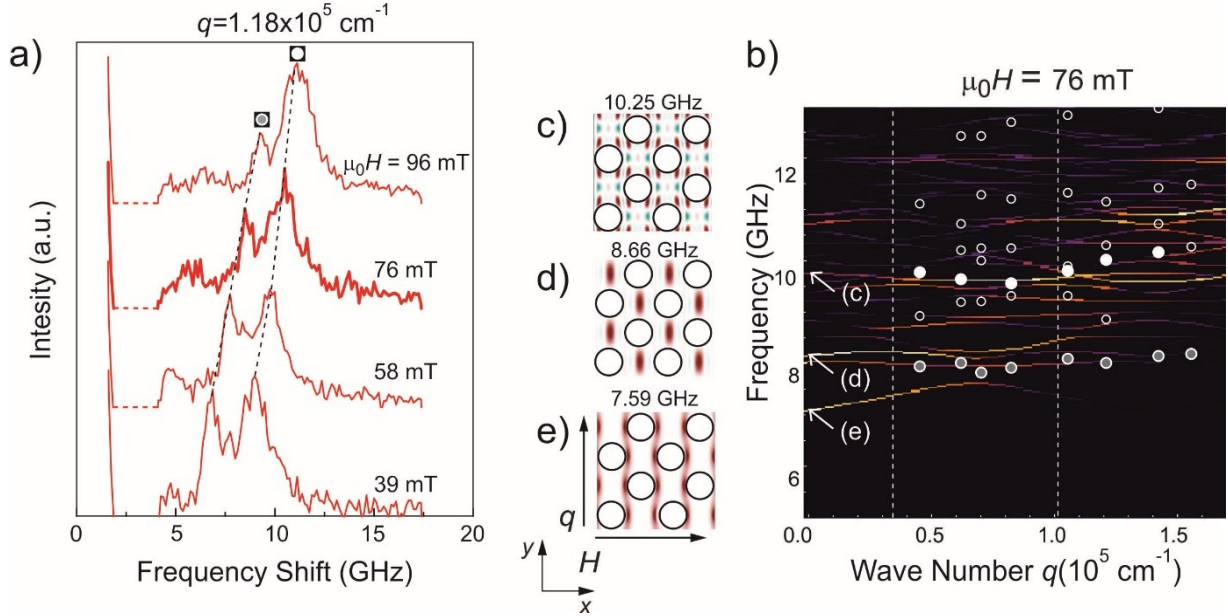

**Fig. 6.** (a) Measured BLS spectra for square ADL corresponding to the different values of the magnetic field applied along the diagonal of the square unit cell ($\phi_H = 0^0$). The spectra were recorded for the incidence angle $\theta = 30^o$ ($q = 1.18 \times 10^{-5}$ cm$^{-1}$) of the laser beam in Damon-Eshbach configuration. (b) BLS intensity map calculated using PWM with BLS experimental points (circles) for Damon-Eshbach configuration in dependence on the wavevector. The filled points denotes the modes of highest BLS intensity. The sections of line colored in light colors correspond to the highest BLS cross-section. The thin vertical lines denotes the borders of the BZs. The white arrows point at the most intensive modes in center of the BZ for which the profiles of out-of-plane component of the dynamical magnetization were plotted in (c), (d), and (e). The color in profiles refers to the phase of dynamical magnetization and the intensity of the color denotes the magnitude of the SW amplitude.

The modes found in PWM at 7.59 and 8.66 GHz at 76 mT (Fig. 6(e) and (d), respectively) are the same as modes obtained in MS shown in Fig. 3(c) and (d), respectively. These modes have largest BLS intensity at small wavenumbers. The SW amplitude distribution of the FM mode suggests, that this mode will be able to propagate with noticeable group velocity. Indeed, the slope of its dispersion branch (i.e. group velocity) is largest in this spectra. Two modes of higher frequency and large simulated BLS intensity have the amplitudes quite concentrated in the areas between next-neighbor antidots, therefore their dispersion branches are flat, which is equivalent to the small values of group velocity for this modes. The measured BLS frequencies corresponding to the modes of highest intensity shows some agreement with the PWM results and show the properties known from the previous studies.[27,51]

Fig. 7(a) presents the BLS spectra for the bi-component ADL at fixed magnitude of the wave number $q = 1.8 \times 10^{-5}$ cm$^{-1}$ and a few values of the external magnetic field (39, 58, 76, 96, 115 mT). Three SWs are clearly identified in the spectra. The measured and calculated SW dispersion relation of this ADL is presented in Fig. 7(b), jointly with the SW profiles obtained from PWM from the center of the first BZ shown in Fig. 7(c-e). Using MS we found two kinds of EMs in this system (see Fig. 4a). Here, the PWM overestimates again the frequency of these EMs. The wells of demagnetizing field around the larger antidots are deeper and wider than for the smaller ones. Therefore we see in PWM calculations the modes which are mixture of the EM of larger antidots and FM (Fig. 7(e)), further on called FM. This mode has similar spatial distribution to the corresponding FM mode in the square ADL (Fig. 6(e)) and similar slope in the first BZ. The frequency of FM in bi-component ADL is slightly lowered in comparison to the frequency of FM mode in square ADL. It can be understandable if we take into account that this mode is concentrated in the vicinity of the wells of demagnetizing field. This lowering is bigger for bi-component ADL where two kinds of such wells exist.

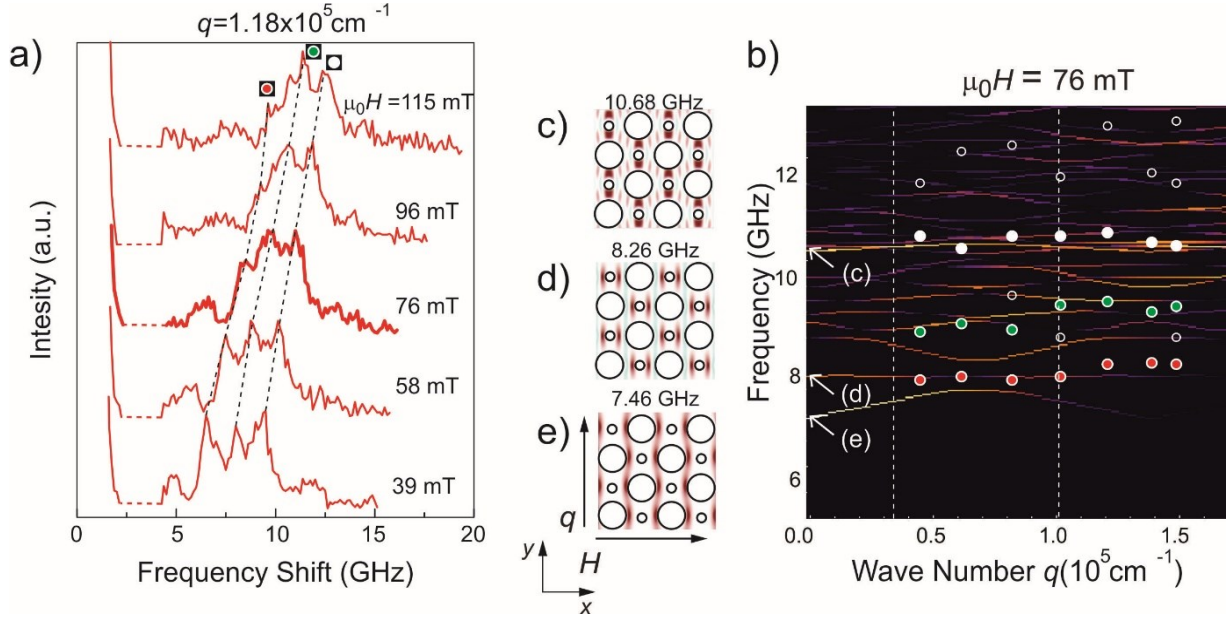

**Fig. 7.** (a) Measured BLS spectra for bi-component ADL corresponding to the different values of the magnetic field applied along the diagonal of the square unit cell ($\phi_H = 0^0$). The spectra were collected for the incidence angle $\theta = 30^o$ ($q = 1.18 \times 10^{-5}$ cm$^{-1}$). (b) BLS cross-section calculated using PWM with BLS experimental points (circles) in Damon-Eshbach configuration where magnetic field was applied in one of the principal directions of the ADL ($\phi=0^0$). The filled points denotes the modes of highest BLS intensity. The sections of line lighter in red correspond to the highest BLS cross-sections. The thin vertical lines denotes the borders of BZs. The white arrows shows the most intensive modes in center of the BZ for which the profiles of out-of-plane component of dynamical magnetization were plotted in (c), (d) and (e). The color in profiles refers to the phase of the SW amplitude and the intensity of color denotes the amplitude.

The next intensive mode is concentrated mostly around smaller antidots, in the wells of the demagnetizing field and therefore we cannot find its counterpart in the square ADL. Due to such localization this mode is weakly dispersive. The high frequency mode of high intensity (at 10.68 GHz) is quite similar in its spatial amplitude distribution to the mode of the square ADL lattice at 8.66 GHz (Fig.7(c)). Both of these modes have the amplitude concentrated in the areas between larger antidots along the *y*-axis, but for bi-component ADL the smaller antidot is placed in the center of this region. This makes area of confinement for the mode smaller in bi-component ADL and shifts up its frequency in reference to the frequency of the corresponding mode in square ADL. This mode is also nondispersive. Overall, for FM and BM good agreement between BLS and PWM results is obtained.

The wave-like structure in PWM calculations (Fig. 8) was simplified as compared to the experimental (Fig. 1(c)) and used in MSs. We omitted in PWM the small deformation remaining from small antidots and also approximated the larger antidots of diamond-like shape (with 275 nm long sides) which is far from the circular antidots used in MSs, but is close to the shapes of the antidots in the sample (see Fig. 1(c)). The width of the air-gaps separating the waveguides was assumed to be 57 nm. These modifications were introduced to use analytical formulas for Fourier coefficients describing spatial distribution material parameters in periodic structure composed of elementary shapes: rectangles and diamonds. This allows to calculate the matrix elements in the PWM eigenvalue problem in Eq. (13) using analytical formulas. For the wave-like structure, we consider the two orthogonal directions of *H* (along the *x* and *y*-axis) and always perpendicular to it SW propagation. For magnetic field applied along the *x*-axis (Fig. 8(a)) the static demagnetizing field is weak and do not affect significantly SW spectrum (e.g. we do not find EM in MS – see Fig.5a). But for magnetic field oriented perpendicular to the wave-like channels (Fig. 8(b)) the strong demagnetizing field shall shift down the frequencies of SWs and generate edge modes (see Fig.5). These EMs are observed in outcomes of PWM only being mixed with FM - see Fig. 8(h), where the intensity of SW amplitude is increased in the vicinity of flat sides of the waveguides perpendicular to the direction of magnetic field. It is also worth to notice that air-gaps brake the exchange interaction between the magnetic wave-like channels. This weakens the strength of coupling for SW propagating along the *y*-axis. It can be expected that SWs along the *x*-axis can more easily propagate through continuous magnetic material. Indeed, this is observed in the calculated and measured dispersion relations shown in Figs. 8(a) and (b). When the wave vector is along the wires (Fig. 8(b)) the group velocity of SWs is larger (0.12 km/s for FM at $q = 0$ – see Fig. 8(h)), while for the propagation across the wave-like channels (Fig. 8(a)) the group velocity is smaller (0.05 km/s for the FM at $q = 0$ – see Fig.8(a)), and comparable for all SW excitations with noticeable BLS intensity.

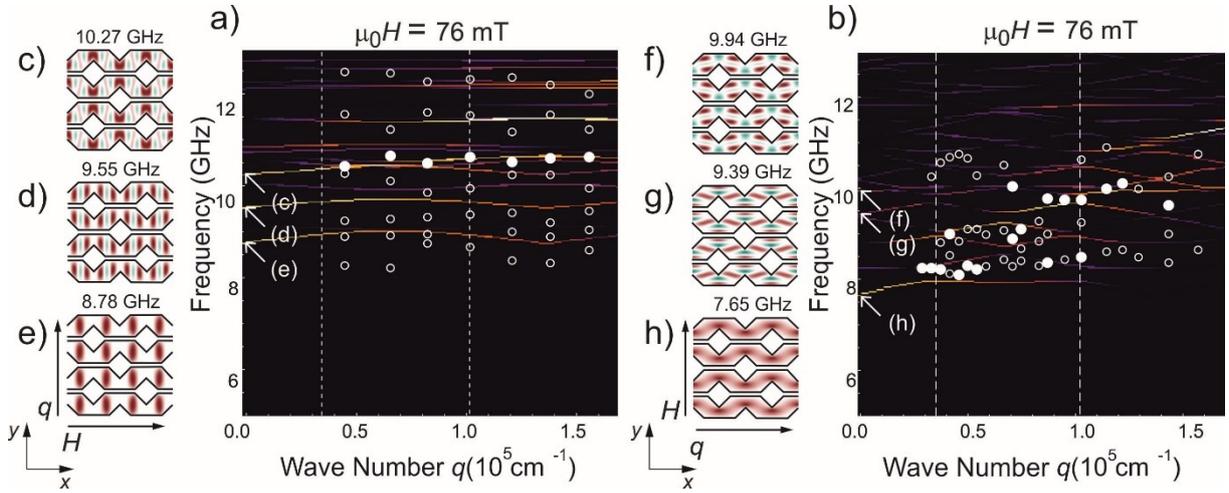

Fig. 8. BLS cross-section calculated using PWM with BLS experimental points (circles) for wave-like ADL in Damon-Eshbach configuration for the in-plane magnetic field oriented at (a) $\phi_H = 0^0$ and (b) $\phi_H = 90^0$. The filled points denote the modes of highest BLS intensity. The sections of line colored in red correspond to the highest calculated BLS cross-section in PWM. The thin vertical lines denotes the borders of the BZs. The white arrows shows the most intensive modes in center of the BZ for which the profiles of the out-of-plane component of dynamical magnetization vector are plotted in (c)-(e) and (f)-(g). The color in profiles refers to the phase of the SW amplitude and the intensity of color denotes the amplitude.

For the magnetic field applied across the wave-like channels (Fig. 8(b)) the FM (mode with significant group velocity for small wave vector) has a continuous amplitude distribution along the channels. The amplitude of the FM for the system presented in Fig. 8(a) is concentrated in the tiled sections of the wave-guides, its profile is close to the FM found in MS (Fig. 5(b)). For this configuration propagation of SW is less effective and this mode have lower group velocity. In considered frequency range we plotted also profiles of two higher modes of significant BLS cross-sections. We found them similar to the spatial distribution of the SWs from MSs in Fig. 5.

# V. Conclusions

We have investigated experimentally (conducting FMR and BLS measurements) the three Py-based thin planar ADLs of square unit cell. We considered the ADLs with unit cells of increasing complexity: simple square ADL, bi-component square ADL and wave-like ADL. The results of the measurements were interpreted with the aid of the micromagnetic simulations and the plane wave method calculations. The good agreement between numerical and experimental results was found. We have found that a small additional antidot in the center of the square unit cell covering only 5% of the area is sufficient to modify the FMR spectra. This small antidot introduces additional edge mode (with amplitude concentrated at edges of the small antidots) and significantly changes the FM dependence on the orientation of the static magnetic field – with maximum (minimum) of the resonance frequency (field) at $\phi_H = 45°$ and $0°$ in square and bi-component ADL, respectively. We have also shown, that two types of defects, irregular shape of anitdots and lattice defects in square ADL can modify noticeably the angular dependence of the SW spectra. The shape distortion from the circular introduce nonequivalence of two perpendicular orientations of the magnetic field. The lattice defects introduce the independences of the FM resonance position on the magnetic field orientation.

Adding air-gaps along parallel lines connecting large and small antidots introduces further interesting changes in the dependence of the FMR spectra on the magnetic field orientation. Here, the fundamental mode continuously transforms into the edge mode when the external magnetic field rotates from the direction parallel to the wave-like channels ($\phi_H = 0°$) to the direction perpendicular to that line. At 30° this mode becomes localized near the edges of the channels but at the same angle the other SW mode becomes most intensive in the FMR spectra. The frequency of this another intensive mode also decrease with further rotation of the magnetic field direction and one another SW excitation becomes most intensive.

The propagation properties of SWs were investigated in the Damon-Eshbach configuration. We demonstrated, that introducing small air-gaps in the wave-like ADL allows to discriminate perpendicular orientation of the magnetic field which were equivalent in the square and bi-component ADLs. When the magnetic field is perpendicular to the wave-like channel the group velocity at the BZ center is high, in the perpendicular orientation it drops to the minimal value.

The acquired in this paper information about SW dynamics in ADLs shall be interesting for the magnonics field and applications of the magnonic crystals in microwave technology, in processing information and also in sensing applications pointed out recently for the ADLs.[52,53]

# Appendix A

**Resonance frequency vs. resonant magnetic field**

In Fig. 10(c) we show the FMR intensity spectra of SWs in the wave-like ADL in dependence on frequency and magnetic field amplitude oriented at $\phi_H = 0°$ obtained from MSs. This result allows to explain the relation between the experimental results obtained in FMR measurements shown in Fig. 2 and MSs results shown in Figs. 3-5. The measurements are performed in dependence on the magnetic field magnitude at fixed frequency 9.4 GHz, marked in Fig. 10(c) with horizontal red-dashed line. The MSs are performed in dependence on frequency at fixed magnetic field magnitude 0.1 T, marked by the vertical black-dashed line in Fig. 10(c). Thus the spectra related to the experimental (Fig. 10(a)) and MSs spectra (Fig. 10(b)) can be obtained by taking cuts of the color map shown in Fig. 10(c) along the horizontal and vertical line, respectively. We see, that FM in measured FMR spectra has largest magnetic field (along red-dashed line), while in MSs it has lowest frequency (along black-dashed line). The BMs are at smaller (higher) fields (frequencies) in measured (simulated) spectra. The EMs doesn't exist here. We decide to make a simulation in frequency domain because this technique requires less computing resources.

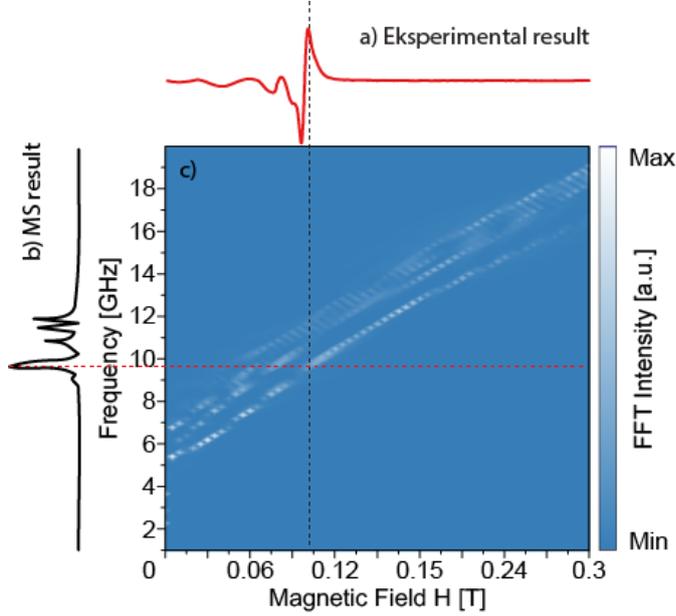

**Fig. 10.** (a) Experimental magnetic field dependent FMR spectra at 9.4 GHz for $\phi_H = 0°$ and (b) frequency dependent FMR spectra obtained from MSs for $\phi_H = 0°$ in static external magnetic field 0.1 T. (c) Color map showing intensity spectra of the SWs in the wavelike ADL obtained from MSs in dependence on frequency and magnetic field amplitude for $\phi_H = 0°$. Cutting of this plot along vertical line (black dashed) gives frequency dependent FMR spectra in (a), cutting along horizontal line (dashed red) gives magnetic field dependent spectra directly related to the measured spectra in (b).

## Appendix B

**Plane wave method for planar 2D magnonic crystals**

In PWM we solve Landau-Lifshitz (LL) equation for the magnetization vector **M** with damping neglected:

$$\frac{d\boldsymbol{M}}{dt} = \gamma\mu_0 \boldsymbol{M} \times \boldsymbol{H}_{eff}, \qquad (1)$$

where $\gamma$ is gyromagnetic ratio, $\mu_0$ is permeability of vacuum, $\mathbf{H}_{\text{eff}}$ is the effective magnetic field. We consider following components of the $\mathbf{H}_{\text{eff}}$: external magnetic field **H**, dynamic exchange field $\mathbf{H}_{\text{ex}}$ and dipolar field $\mathbf{H}_{\text{d}}$, thus $\mathbf{H}_{\text{eff}} = \mathbf{H} + \mathbf{H}_{\text{ex}} + \mathbf{H}_{\text{d}}$. We limit our investigations to linear regime of magnetization dynamics where we could study SW dynamics characterized by harmonic oscillations $e^{i\omega t}$ ($\omega$ is the angular eigenfrequency of SW) on the background of the static magnetic configuration in saturation state. Linearized LL Eq. (1) can be written in the form of two differential equations for complex amplitudes $m_x$ and $m_y$ of the dynamical components of the magnetization vector $\mathbf{m}(\mathbf{r}, t) = [m_x(\mathbf{r})e^{i\omega t}, m_y(\mathbf{r})e^{i\omega t}, 0]$:

$$i\Omega m_x(\mathbf{r}) = Hm_y(\mathbf{r}) - M_S(\mathbf{r})h_{\text{ex},y}(\mathbf{r}) - M_S(\mathbf{r})h_{\text{d},y}(\mathbf{r}) + m_y(\mathbf{r})H_{\text{d},z}(\mathbf{r}), \tag{2}$$

$$i\Omega m_y(\mathbf{r}) = -Hm_x(\mathbf{r}) + M_S(\mathbf{r})h_{\text{ex},x}(\mathbf{r}) + M_S(\mathbf{r})h_{\text{d},x}(\mathbf{r}) - m_x(\mathbf{r})H_{\text{d},z}(\mathbf{r}), \tag{3}$$

where $\Omega = \frac{\omega}{\gamma\mu_0}$. We assume external field parallel to the z-axis $\mathbf{H} = [0,0,H]$ in this derivation. Exchange field in saturation state has only non zero dynamical components: $\mathbf{H}_{\text{ex}}(\mathbf{r},t) = [h_{\text{ex},x}(\mathbf{r})e^{i\omega t}, h_{\text{ex},y}(\mathbf{r})e^{i\omega t}, 0]$. The demagnetizing filed vector can be written in the following form: $\mathbf{H}_{\text{d}}(\mathbf{r},t) = [h_{\text{d},x}(\mathbf{r})e^{i\omega t}, h_{\text{d},y}(\mathbf{r})e^{i\omega t}, H_{\text{d},z}(\mathbf{r})]$, where $h_{\text{d},x}$ and $h_{\text{d},y}$ are dynamical components of demagnetizing field, and $H_{\text{d},z}(\mathbf{r})$ is the static demagnetizing field.

The exchange field in MCs shall include the spatially dependent exchange length $\lambda_{\text{ex}}(\mathbf{r})$ ($\lambda_{\text{ex}}(\mathbf{r}) = 2A(\mathbf{r})/\mu_0 M_S^2(\mathbf{r})$):[43]

$$\mathbf{H}_{\text{ex}}(\mathbf{r},t) = \nabla\lambda_{\text{ex}}^2(\mathbf{r})\nabla\mathbf{m}(\mathbf{r},t). \tag{4}$$

For thin slab of MC and limiting study to the SWs of low frequency, the dynamical magnetization can be assumed uniform across its thickness. Within this assumption, the exact form of x- and y- components of the dynamical exchange field are:

$$h_{\text{ex},\alpha}(\mathbf{r}) = \left[\frac{\partial}{\partial y}\lambda_{\text{ex}}^2(\mathbf{r})\frac{\partial}{\partial y} + \frac{\partial}{\partial z}\lambda_{\text{ex}}^2(\mathbf{r})\frac{\partial}{\partial z}\right]m_\alpha(\mathbf{r}), \tag{5}$$

where $\alpha = x,y$.

The static and dynamical components of the demagnetizing file in a planar periodic structures can be calculated by solving Maxwell equations in magnetostatic approximation with use of electromagnetic boundary conditions at the slab surfaces. The formulas for the static components of the demagnetizing field in the periodic magnetic layer were derived by Kaczer and Martinowa[54] and extended to the dynamical components in Ref. [55]:

$$h_{\text{d},x}(\mathbf{r}) = \tag{6}$$
$$-\sum_{\mathbf{G}} m_x(\mathbf{G})C(|\mathbf{q}+\mathbf{G}|,x)e^{-i(\mathbf{G}+\mathbf{q})\cdot\mathbf{r}_\parallel} + i\frac{q_y+G_y}{|q_y+G_y|}m_y(\mathbf{G})S(|\mathbf{q}+\mathbf{G}|,x)e^{-i(\mathbf{G}+\mathbf{q})\cdot\mathbf{r}_\parallel},$$

$$h_{\text{d},y}(\mathbf{r}) = \tag{7}$$
$$-\sum_{\mathbf{G}} i\frac{q_y+G_y}{|q_y+G_y|}m_x(\mathbf{G})S(|\mathbf{q}+\mathbf{G}|,x)e^{-i(\mathbf{G}+\mathbf{q})\cdot\mathbf{r}_\parallel} + \frac{(q_y+G_y)^2}{|\mathbf{q}+\mathbf{G}|^2}m_y(\mathbf{G})$$
$$[1-C(|\mathbf{q}+\mathbf{G}|,x)]e^{-i(\mathbf{G}+\mathbf{q})\cdot\mathbf{r}_\parallel},$$

$$H_{\text{d},z}(\mathbf{r}) = -\sum_{\mathbf{G}}\frac{G_z^2}{G^2}M_S(\mathbf{G})[1-C(|\mathbf{G}|,x)]e^{-i\mathbf{G}\cdot\mathbf{r}_\parallel}, \tag{8}$$

where $\mathbf{G}$ denotes the reciprocal lattice vectors of the MC (for the square lattice $\mathbf{G} = \frac{2\pi}{a}(n_y,n_z)$ where $n_y$ and $n_z$ are integers) and $\mathbf{r}_\parallel = [0,y,z]$ is the in-plane position vector. $m_\alpha(\mathbf{G})$, are the coefficients of the Fourier expansions for dynamical components of the magnetization, fulfilled Bloch theorem:

$$m_\alpha(\mathbf{r}) = \sum_{\mathbf{G}} m_\alpha(\mathbf{G})e^{-i(\mathbf{G}+\mathbf{q})\cdot\mathbf{r}_\parallel}. \tag{9}$$

The functions: $C(q,x)$ and $S(q,x)$ are defined as:

$$C(q,x) = \frac{\cosh(qx)}{\sinh(qx)+\cosh(qx)}, \quad S(q,x) = \frac{\sinh(qx)}{\sinh(qx)+\cosh(qx)} \qquad (10)$$

In MCs the material parameters $M_S(\mathbf{r})$ and $\lambda_{ex}^2(\mathbf{r})$ are periodic functions in space and can be expanded in Fourier series:

$$M_S(\mathbf{r}) = \sum_{\mathbf{G}} M_S(\mathbf{G})e^{-i\mathbf{G}\cdot\mathbf{r}_\parallel}, \quad \lambda_{ex}^2(\mathbf{r}) = \sum_{\mathbf{G}} \lambda_{ex}^2(\mathbf{G})e^{-i\mathbf{G}\cdot\mathbf{r}_\parallel}. \qquad (11)$$

The exchange field in Eq. (5) with using Eqs. (9) and (11) can be written as Fourier series for dynamical magnetization:

$$h_{ex,\alpha}(\mathbf{r}) = -\sum_{\mathbf{G},\mathbf{G}'}(\mathbf{q}+\mathbf{G})\cdot(\mathbf{q}+\mathbf{G}')\lambda_{ex}^2(\mathbf{G}'-\mathbf{G})m_\alpha(\mathbf{G})e^{-i(\mathbf{G}+\mathbf{q})\cdot\mathbf{r}_\parallel}. \qquad (12)$$

Substituting Eq. (6)-(8) and (12) to Eqs. (2)-(3) we obtain the eigenvalue problem for the reduced frequencies $\Omega$ of the SW eigenmodes:

$$\widehat{\mathbf{M}}\, \mathbf{m_G} = \Omega\, \mathbf{m_G}, \qquad (13)$$

where $\mathbf{m_G} = [\ldots m_x(\mathbf{G_i})\ldots,\ldots m_y(\mathbf{G_i})\ldots]^T$ is the vector of Fourier coefficients for the $x$ and $y$ components of the dynamical magnetization. The matrix $\widehat{\mathbf{M}}$ is a block matrix:

$$\widehat{\mathbf{M}} = \begin{pmatrix} M_{xx} & M_{xy} \\ M_{yx} & M_{yy} \end{pmatrix}, \qquad (14)$$

where:

$$(M_{xx})_{i,j} = -i\frac{q_y+G_{y,j}}{|\mathbf{q}+\mathbf{G}_j|}M_S(\mathbf{G_i}-\mathbf{G_j})S(\mathbf{q}+\mathbf{G}_j,x) = -(M_{yy})_{i,j}, \qquad (15)$$

$$(M_{xy})_{i,j} = H_0 - \sum_l(\mathbf{q}+\mathbf{G_j})\cdot(\mathbf{q}+\mathbf{G_l})\,\lambda_{ex}^2(\mathbf{G_l}-\mathbf{G_j})M_S(\mathbf{G_i}-\mathbf{G_l}) \qquad (16)$$

$$+\frac{(q_y+G_{y,j})^2}{|\mathbf{q}+\mathbf{G}_j|^2}M_S(\mathbf{G_i}-\mathbf{G_j})\left(1-S(\mathbf{q}+\mathbf{G}_j,x)\right)$$

$$-\frac{(G_{z,i}+G_{z,j})^2}{|\mathbf{G}_i+\mathbf{G}_j|^2}M_S(\mathbf{G_i}-\mathbf{G_j})\left(1-S(\mathbf{G}_i+\mathbf{G}_j,x)\right),$$

$$(M_{xy})_{i,j} = -H + \sum_l(\mathbf{q}+\mathbf{G_j})\cdot(\mathbf{q}+\mathbf{G_l})\,\lambda_{ex}^2(\mathbf{G_l}-\mathbf{G_j})M_S(\mathbf{G_i}-\mathbf{G_l}) \qquad (17)$$

$$-M_S(\mathbf{G_i}-\mathbf{G_j})C(\mathbf{q}+\mathbf{G}_j,x)$$

$$+\frac{(G_{z,i}+G_{z,j})^2}{|\mathbf{G}_i+\mathbf{G}_j|^2}M_S(\mathbf{G_i}-\mathbf{G_j})\left(1-S(\mathbf{G}_i+\mathbf{G}_j,x)\right).$$

The dispersion relation of SWs and SW mode profiles can be obtained from the eigenvalues $\Omega$ and eigenvectors $\mathbf{m_G}$ of the Eq. (11). The Eq. (11) can be solved numerically by taking the finite number of reciprocal lattice vectors in the Fourier expansions of $m_\alpha(\mathbf{r})$, $M_S(\mathbf{r})$ and $\lambda_{ex}^2(\mathbf{r})$ in Eqs. (9) and (11). The Fourier coefficients $M_S(\mathbf{G})$ and $\lambda_{ex}^2(\mathbf{G})$ for the shapes of the unit cells considered in this paper can be calculated analytically. For antidots or inclusions of circular or square shape, they can be found in Ref. [56].

The special care need to be taken for the nonmagnetic parts of the ADL, which in the presented formulation of the PWM need to be included into computation. It can be done by introducing artificial magnetic materials (with the low magnetization saturation and low value of the exchange length) instead of the air. As demonstrated in Ref. [57] this approach gives correct results for FM and BM, however overestimates frequencies of the EMs. This is because artificial material introduces pinning of the magnetization at the antidots edges,[58] which shifts the frequencies of the EMs to higher values.

## Acknowledgments

This research is supported by the SYMPHONY project operated within the Foundation for Polish Science within the Team Programme co-financed by the EU European Regional Development Fund, Grant No. OPIE 2007-2013, and partially received funding from Polish National Science Centre Project No. UMO-2012/07/E/ST3/00538and from the European Union Horizon2020 research and innovation programme under the Marie Sklodowska-Curie grant agreement No. 644348 (MagIC). The simulations were partially performed at Poznan Supercomputing and Networking Center (grant No 209).